\documentstyle[aps,prl,floats,preprint,epsfig]{revtex}

\begin{document}

\renewcommand{\topfraction}{1.0}
\renewcommand{\bottomfraction}{1.0}
\renewcommand{\textfraction}{0.0}
\renewcommand{\labelitemii}{ }

\def\Uone {$\Upsilon(1S)$} 
\def\Utwo {$\Upsilon(2S)$} 
\def\Uthree {$\Upsilon(3S)$}
\def\Ufour {$\Upsilon(4S)$}
\def\Ecm  {E_{cm}}

\def\decfoursonespipi {$\Upsilon(4S)\rightarrow \Upsilon(1S)\pi^+\pi^-$}
\def\decfourstwospipi {$\Upsilon(4S)\rightarrow \Upsilon(2S)\pi^+\pi^-$}
\def\decthreesonespipi {$\Upsilon(3S)\rightarrow \Upsilon(1S)\pi^+\pi^-$}
\def\dectwosonespipi {$\Upsilon(2S)\rightarrow \Upsilon(1S)\pi^+\pi^-$}
\def\decthreestwospipi {$\Upsilon(3S)\rightarrow \Upsilon(2S)\pi^+\pi^-$}
\def\proeethreesgam {$e^+e^- \rightarrow \Upsilon(3S)\gamma$}
\def\proeetwosgam {$e^+e^- \rightarrow \Upsilon(2S)\gamma$}
\def\decnsmspipi {$\Upsilon(nS)\rightarrow \Upsilon(mS)\pi^+\pi^-$}
\def\decmsll {$\Upsilon(mS)\rightarrow e^+e^-,\mu^+\mu^-$}

 \def\xsecuthreeon {  17.5\pm   3.0\pm   1.6 }
 \def\xsecuthreeof {  27.3\pm   5.0\pm   2.5 }
 \def\xsecutwoon {  15.5\pm   1.3\pm   1.3 }
 \def\xsecutwoof {  16.3\pm   1.8\pm   1.3 }
 \def\fitdf { 22 }
 \def\lumtotal { 4.19 }
 \def\lumon { 2.74 }
 \def\lumoff { 1.45 }
 \def\ulfoursones {   1.2 }
 \def\ulfourstwos {   3.9 }

\preprint{\tighten\vbox{ \hbox{\hfil CLNS 98/1576}
                         \hbox{\hfil CLEO 98-12}
}}

\title{$\Upsilon$ dipion transitions at energies near the \Ufour.\\}

\author{CLEO Collaboration}
\date{\today}
\maketitle
\tighten

\begin{abstract}
Using a \lumtotal\ fb$^{-1}$ data sample collected with the CLEO~II detector at the Cornell Electron Storage Ring, we have searched for dipion transitions between pairs of $\Upsilon$ resonances at center of mass energies $\Ecm=10.58$ GeV and $\Ecm=10.52$ GeV. We obtain the 90\% confidence level upper limits ${\cal B}(\Upsilon(4S)\rightarrow \Upsilon(2S)\pi^+\pi^-) < \ulfourstwos \times 10^{-4}$ and ${\cal B}(\Upsilon(4S)\rightarrow \Upsilon(1S)\pi^+\pi^-) < \ulfoursones \times 10^{-4}$. We also observe the transitions $\Upsilon(3S)\rightarrow \Upsilon(1S)\pi^+\pi^-$, $\Upsilon(3S)\rightarrow \Upsilon(2S)\pi^+\pi^-$, and $\Upsilon(2S)\rightarrow \Upsilon(1S)\pi^+\pi^-$, from which we measure the cross-sections for the radiative processes $e^+e^- \rightarrow \Upsilon(3S)\gamma$ and $e^+e^- \rightarrow \Upsilon(2S)\gamma$. We obtain $\sigma_{ee \rightarrow \Upsilon(3S)\gamma}=(\xsecuthreeon)$~pb and $\sigma_{ee\rightarrow \Upsilon(2S)\gamma}=(\xsecutwoon)$~pb at $\Ecm=10.58$~GeV, and $\sigma_{ee \rightarrow \Upsilon(3S)\gamma}=(\xsecuthreeof)$~pb and $\sigma_{ee\rightarrow \Upsilon(2S)\gamma}=(\xsecutwoof)$~pb at $\Ecm=10.52$~GeV, which we compare with theoretical predictions.
\end{abstract}

\newpage
{
\renewcommand{\thefootnote}{\fnsymbol{footnote}}
\begin{center}
S.~Glenn,$^{1}$ Y.~Kwon,$^{1,}$%
\footnote{Permanent address: Yonsei University, Seoul 120-749, Korea.}
A.L.~Lyon,$^{1}$ S.~Roberts,$^{1}$ E.~H.~Thorndike,$^{1}$
C.~P.~Jessop,$^{2}$ K.~Lingel,$^{2}$ H.~Marsiske,$^{2}$
M.~L.~Perl,$^{2}$ V.~Savinov,$^{2}$
% ...
\end{center}
 
\small
\begin{center}
$^{1}${University of Rochester, Rochester, New York 14627}\\
$^{2}${Stanford Linear Accelerator Center, Stanford University, Stanford,
California 94309}\\
% ...
\end{center}

\setcounter{footnote}{0}
}
\newpage

\section{Introduction}

Bottomonium dipion transitions have been the subject of many studies~\cite{review,Yan,ZhouKuang}. So far, theoretical efforts have concentrated on investigating dipion transitions between pairs of $\Upsilon$ resonances below $B{\overline B}$ threshold production, in part because of complexities in the theoretical analysis of coupled-channel effects above $B{\overline B}$ threshold. There are no experimental results on $\Upsilon(4S)$ dipion transitions; one would generally expect very small branching fractions for such $\Upsilon(4S)$ decays due to the large OZI-allowed width for $\Upsilon(4S)\rightarrow B\overline{B}$. Nevertheless the large amount of CLEO~II \Ufour\ resonance data and our familiarity with the systematics of such transitions~\cite{u2trans} make it worthwhile to perform a dedicated search for the transitions $\Upsilon(4S)\rightarrow \Upsilon(2S)\pi^+\pi^-$ and $\Upsilon(4S)\rightarrow \Upsilon(1S)\pi^+\pi^-$. 

We can also measure the cross-sections for the processes \proeethreesgam\ and \proeetwosgam\ by reconstructing the decay chains \decnsmspipi,\decmsll\ with $(n,m)=(3,1),(2,1),(3,2)$. These processes are important in determining the accuracy of theoretical calculations~\cite{Rtheory} and experimental measurements~\cite{Rexper} of the cross-section for hadron production in $e^+e^-$ annihilations at $\sqrt{s}\approx 10$ GeV. The process $e^+e^-\rightarrow \Upsilon(nS)\gamma \rightarrow\ \gamma + hadrons$ comprises one of the largest systematic uncertainties to the measurement of $R=\sigma(e^+e^-\rightarrow hadrons)/\sigma(e^+e^-\rightarrow \mu^+\mu^-)$ in the $\Upsilon$ region.

\section{Detector}

The CLEO~II detector, described in detail elsewhere~\cite{cleodetector}, is a general-purpose magnetic spectrometer and calorimeter for measuring charged and neutral particles. The major subsystems of the detector (in order of increasing radius from the beam pipe) are the central detector, time-of-flight scintillators, the crystal calorimeter, 1.5-Tesla superconducting coil, and the muon chambers and magnetic flux return units. The central detector consists of three concentric drift chambers and is used for reconstruction of charged particle momenta and measurements of specific ionization energy loss ($dE/dx$) for particle identification. This system achieves a momentum resolution $(\delta p/p)^2=(0.0015p)^2 + (0.005)^2$, where $p$ is the momentum in GeV/$c$, and covers 95\% of the solid angle. The time-of-flight system is used in two ways: for the lower-level trigger, and for measuring the flight time of particles to help in particle identification.
The crystal calorimeter, which measures the energies deposited by neutral and charged particles, consists of 7800 thallium-doped cesium iodide (CsI) crystals arranged in a barrel and two endcaps. The central barrel region of the calorimeter covers 75\% of the solid angle and achieves an energy resolution $\delta E/E(\%)=0.35/E^{0.75} + 1.9 -0.1E$, where $E$ is the shower energy in GeV. The endcaps extend the solid angle coverage to about 95\% of $4\pi$, although they provide poorer energy resolution than the barrel region. The muon identification system, also arranged as an octagonal barrel and two endcaps, uses proportional tracking chambers for muon detection. These chambers are sandwiched between and behind the iron slabs that provide the magnetic field flux return. 

In our analysis the JETSET~\cite{jetset} program is used as a Monte Carlo event generator. We use a GEANT~\cite{geant} based detector simulation package to propagate and decay the final state particles in the CLEO~II detector.

\section{Event selection}

In our analysis of \decnsmspipi\ transitions\footnote{$(n,m)=(4,1),(4,2),(3,1),(2,1),(3,2)$} we reconstruct the $\Upsilon(mS)$ exclusively from the decays \decmsll. The following selection criteria are common to all five transitions: (1) we require a total of four good quality primary charged tracks in the event with zero net charge; (2) two of them (the lepton candidates) must have momenta greater than 3.5 GeV/$c$ and originate from the interaction region, defined as a cylindrical volume of 3 mm radius and 10 cm length aligned along the beam axis and centered on the $e^+e^-$ collision point; (3) the other two tracks (the pion candidates) must have momenta less than 1 GeV/$c$ and originate from a similar cylindrical volume 4 mm $\times$ 12 cm centered on the interaction point; (4) we identify electrons by requiring the ratio of the associated electromagnetic shower energy deposited in the calorimeter to the momentum of the matching track to be close to unity and the lateral pattern of energy deposition to be consistent with the electron hypothesis; (5) muons are identified by requiring the maximum penetration depth of the muon track candidate into the muon system absorber to be greater than three hadronic absorption lengths; (6) we require the cosine of the opening angle between the pion tracks to satisfy $\cos(p_{\pi}p_{\pi})<0.9$ to suppress background from $e^+e^-\rightarrow e^+e^-\gamma$ events with $\gamma$-conversion when the resulting $e^+e^-$ pair fakes a $\pi^+\pi^-$ pair; and (7) to further reduce this background in the $ee$ channel (only) we require that at least one of the pion candidates must have its specific ionization energy loss measurement ($dE/dx$) consistent with the pion hypothesis.\footnote{We make no requirement on the dilepton invariant mass, because it has very small effect on the signal to background ratio while reducing the signal yields.}

\begin{figure}[htb]
\center
\epsfig{file=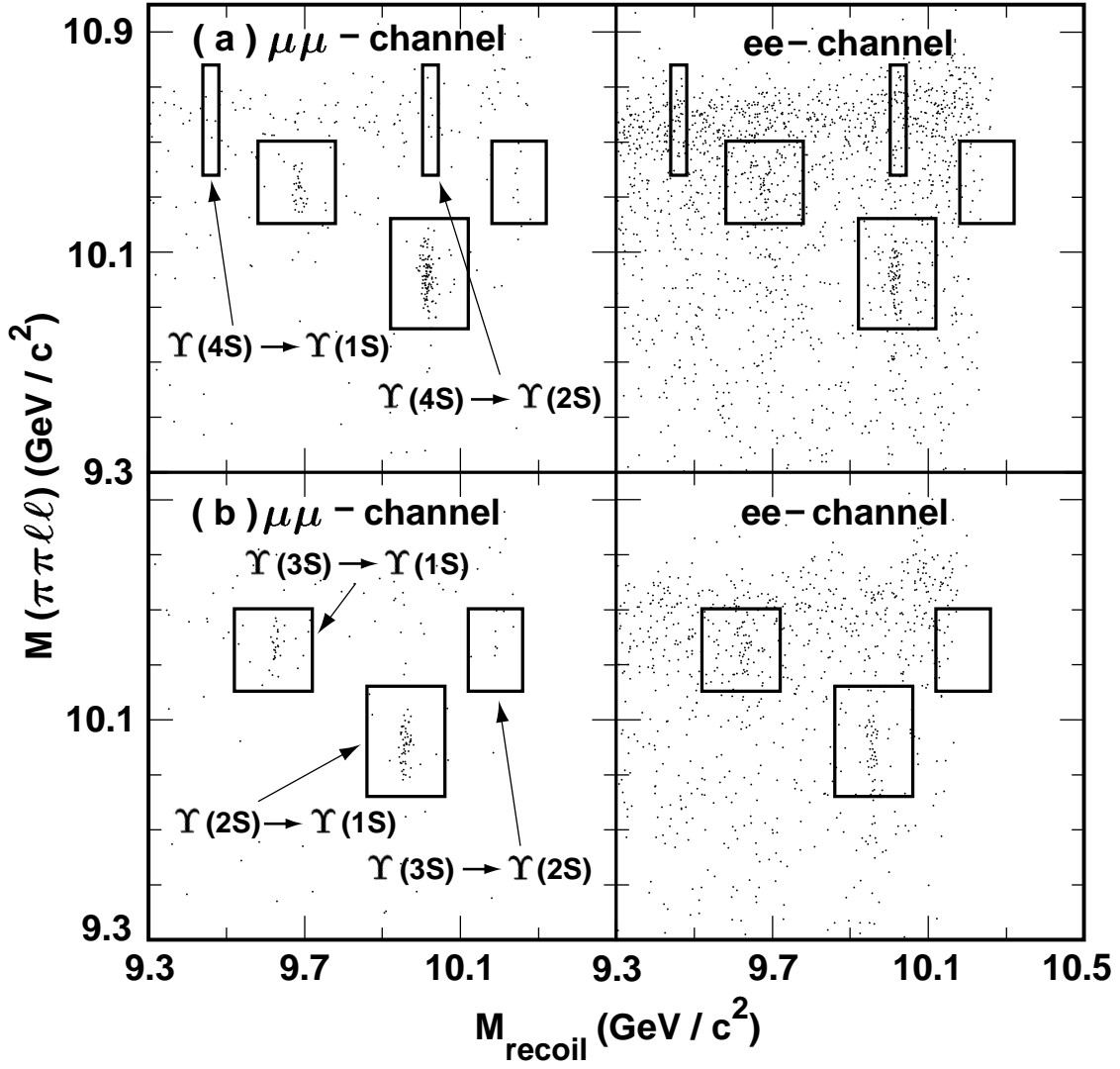, width=15cm}
\caption[]{Plots of $\pi\pi ll$ invariant mass vs recoil mass for the \decnsmspipi\ transitions from data: a) at $\Ecm=10.58$ GeV (on the $\Upsilon(4S)$), b) at $\Ecm=10.52$ GeV (below the $\Upsilon(4S)$).}\label{fig:g2pi2ldtscat}
\end{figure}

We search for a signal from the transitions of interest by plotting the invariant mass of the $\pi^+\pi^-l^+l^-$ system vs the mass recoiling against the dipion system $M_{recoil}=\sqrt{(\Ecm-\Sigma E_{\pi})^{2}-(\Sigma{\bf p}_{\pi})^{2}}$ as shown in Figure~\ref{fig:g2pi2ldtscat}: the upper plots are for $\Ecm=10.58$ GeV and the lower plots are for $\Ecm=10.52$ GeV (the boxes denote our signal regions). Peaks from the $\Upsilon(3S) \rightarrow \Upsilon(1S)$  and $\Upsilon(2S) \rightarrow \Upsilon(1S)$ transitions are clearly seen in all four distributions. One can also notice a much smaller signal from the $\Upsilon(3S) \rightarrow \Upsilon(2S)$ transition in the $\mu\mu$ channel. This signal is not seen in the $ee$ channel because of a cutoff in $M_{recoil}$ due to the absence of $dE/dx$ information for tracks with momentum less than \mbox{$\sim 100$ MeV/$c$}; such tracks do not reach into the outer drift chamber where the $dE/dx$ measurement is performed. 

Because the transitions $\Upsilon(3S) \rightarrow \Upsilon(2S)$, $\Upsilon(3S) \rightarrow \Upsilon(1S)$ and $\Upsilon(2S) \rightarrow \Upsilon(1S)$ occur in events with initial state radiation, the peak of the recoil mass for these transitions is shifted from the mass value of the daughter $\Upsilon$ by roughly $E_{\gamma}$, the energy of the unobserved initial state radiation photon (Table~\ref{tab:enshift}). 

\begin{table}[hbt]
\center
\caption[]{\small The initial state radiation photon energies and the recoil mass peak positions for $\Ecm=10.58(10.52)$ GeV.}\label{tab:enshift}
\begin{tabular}{ccc}
 Transition  & $E_{\gamma}$ (GeV) & $M_{recoil}^{peak}$ (GeV) \\ \hline
$\Upsilon(3S)\rightarrow \Upsilon(1S)$ & 0.23(0.17) & 9.67(9.61) \\
$\Upsilon(2S)\rightarrow \Upsilon(1S)$   & 0.56(0.50) & 10.02(9.96) \\
$\Upsilon(3S)\rightarrow \Upsilon(2S)$ & 0.23(0.17) & 10.25(10.19) \\
\end{tabular}
\end{table}

\clearpage

\section{Search for the transitions \boldmath{\decfourstwospipi} and \boldmath{\decfoursonespipi}}

\subsection{Extraction of upper limits}

\begin{figure}[htb]
\center
\epsfig{file=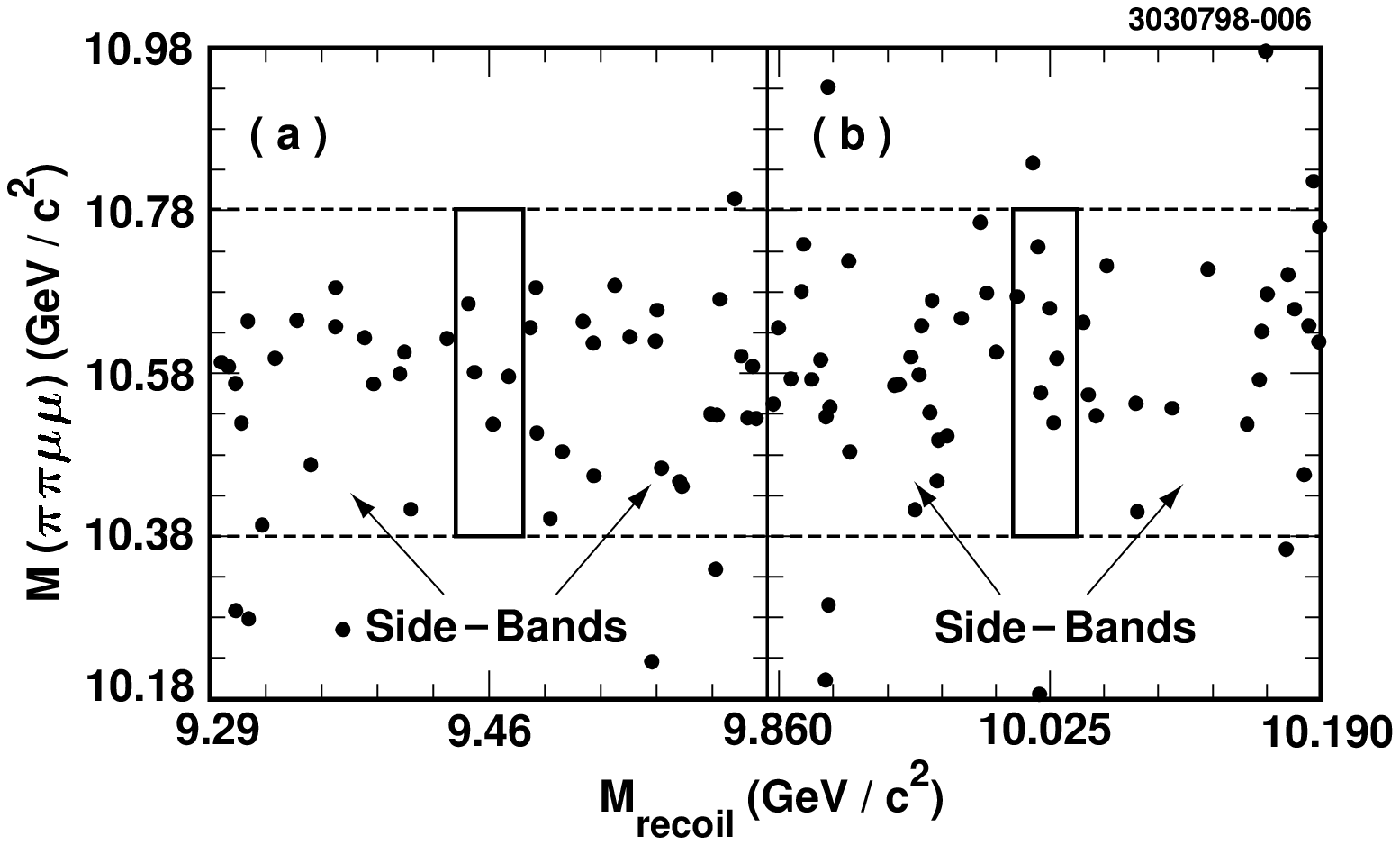}
\caption[]{The $\pi\pi\mu\mu$ invariant mass vs $\pi^+\pi^-$ recoil mass  for a) \decfoursonespipi\, and b) \decfourstwospipi\ at $\Ecm=10.58$ GeV.}
\label{fig:4sallmuscat}
\end{figure}

As one can see from Figure~\ref{fig:g2pi2ldtscat}, although there are data points in the signal regions\footnote{The size of the signal region is defined as a box $\pm 3$ standard deviations wide in each of the variables $M_{recoil}$ and $m_{\pi\pi ll}$. This numerically corresponds to (9.44,9.48) $\cap$ (10.38,10.78) for $\Upsilon(4S) \rightarrow \Upsilon(1S)\pi\pi$ and (10.003,10.043) $\cap$ (10.38,10.78) for $\Upsilon(4S) \rightarrow \Upsilon(2S)\pi\pi$.} for \decfourstwospipi\ and \decfoursonespipi, there is no apparent clustering of the signal. Because of the overwhelmingly large backgrounds in the $ee$ channel, we limit our analysis of these two transitions to the $\mu\mu$ channel. In Figure~\ref{fig:4sallmuscat} close-ups of the signal regions for the \Ufour\ dipion transitions are shown. We employ a ``grand side-band'' technique to evaluate the background: we count the events in the side-bands\footnote{The side-bands are the horizontal strips of dimensions (9.29,9.63) $\cap$ (10.38,10.78) for $\Upsilon(4S) \rightarrow \Upsilon(1S)\pi\pi$ and (9.853,10.193) $\cap$ (10.38,10.78) for $\Upsilon(4S) \rightarrow \Upsilon(2S)\pi\pi$ in the variables $M_{recoil}$ and $m_{\pi\pi ll}$ respectively, excluding the signal regions.} and extrapolate the background event yield into the signal region. Numbers of observed events and numbers of expected background events are reported in Table~\ref{tab:4strans}, along with upper limits (at 90\% confidence level) on the number of signal events. Also shown are the efficiencies calculated from Monte Carlo simulation. As a cross-check we performed the scaled continuum subtraction to estimate the background, which yields consistent results, given the limited continuum statistics.

\begin{table}[htb]
\center
\caption[]{\small Efficiencies, numbers of events, branching fractions and rates for the charged \Ufour\ dipion transitions observed in the $\mu\mu$ channel.}\label{tab:4strans}
\begin{tabular}{ccccccc}
 Transition  & $\epsilon$ (\%) & $N^{observed}$ & $N^{background}_{expected}$ & $N^{signal}_{upper\ limit}$ & ${\cal B}\ (\times 10^{-4})$ & $\Gamma$\ (keV) \\ \hline
 $\Upsilon(4S)\rightarrow \Upsilon(1S)$ & $ 48.6\pm   1.6$&   4 &   5.2 &   4.3 & $<  1.2$ & $<  2.5$ \\
 $\Upsilon(4S)\rightarrow \Upsilon(2S)$ & $ 38.3\pm   1.3$ &   6 &   5.9&   5.6 & $<  3.9$ & $<  8.2$ \\
\end{tabular}
\end{table}

We calculate upper limits on the branching fractions and partial widths for $\Upsilon(4S)\rightarrow \Upsilon(mS)\pi^+\pi^-$ using the formula ${\cal B}=N^{signal}_{upper\ limit}/(\epsilon {\cal B}_{\mu\mu} \sigma {\cal L})$, where ${\cal B}_{\mu\mu}$ is the $\Upsilon(mS)$ muonic branching fraction taken from~\cite{pdgupsilon} (${\cal B}_{\mu\mu}=(2.48\pm 0.07)\%$ for \Uone\ and ${\cal B}_{\mu\mu}=(1.31\pm 0.21)\%$ for \Utwo), $\sigma=(1.074\pm 0.020)$ nb is the average \Ufour\ production cross-section at CESR, and ${\cal L}=(\lumon\pm 0.02)$ fb$^{-1}$ is the integrated luminosity of our on-resonance data sample. Upper limits resulting from these calculations are:\footnote{We follow the procedure described in the Review of Particle Physics~\cite{poisson}, and incorporate systematic uncertainties according to~\cite{syserinuplim}.}
\[{\cal B}(\Upsilon(4S)\rightarrow \Upsilon(1S)\pi^+\pi^-) < \ulfoursones \times 10^{-4}\ (C.L.=90\%) \]
\[{\cal B}(\Upsilon(4S)\rightarrow \Upsilon(2S)\pi^+\pi^-) < \ulfourstwos \times 10^{-4}\ (C.L.=90\%) \]

\subsection{Systematic errors}

The dominant systematic errors in our search for the \Ufour\ dipion transitions are due to uncertainties in the \Uone\ and \Utwo\ muonic branching fractions and in the track-finding efficiency. Other large sources of systematic errors include trigger efficiency uncertainties and the uncertainty in the \Ufour\ production cross-section. The complete breakdown of systematic errors is given in Table~\ref{tab:4ssysterrors} (relative errors in percent).  

\begin{table}[htb]
\center
\caption[]{\small Sources and magnitudes of systematic errors in $\Upsilon(4S)\rightarrow \Upsilon(mS)\pi\pi$ transitions.}\label{tab:4ssysterrors}
\begin{tabular}{lcc}
        &\multicolumn{2}{c}{Systematic error (\%)} \\ \cline{2-3}
Source &\decfoursonespipi &\decfourstwospipi \\ \hline
Tracking                         & 2.8 & 2.8 \\ 
Finite MC sample                 & 1.0 & 1.0 \\
Trigger efficiency               & 1.5 & 1.5 \\
\Ufour\ production cross-section & 2.0 & 2.0 \\
Luminosity                       & 0.9 & 0.9 \\
$\Upsilon$ muonic branching ratio    & 2.8 & 16.0 \\ \hline
Total                            & 4.9 & 16.5 \\
\end{tabular}
\end{table}

\subsection{Discussion and conclusions}

There are several predictions for the rates of dipion transition between heavy quarkonia states~\cite{Yan,ZhouKuang} below $B\overline{B}$ threshold. Naively, we might expect some suppression for \Ufour\ dipion transitions compared with the corresponding \Uthree\ transitions because of an additional node in the \Ufour\ wave function; this is compensated, to some extent, by the larger available phase space in the \Ufour\ transition. Unfortunately, the proximity of the \Ufour\ resonance to the $B\overline{B}$ threshold leads to the necessity of estimating  coupled-channel contributions to the transition rates. Although there exists a model for calculating coupled-channel effects from the virtual process $\Upsilon\rightarrow B\overline{B}\rightarrow \Upsilon'$~\cite{ZhouKuang}, there is no such model for the real mixing of $\Upsilon$ and $B\overline{B}$ states.

Our measured upper limits on the branching fraction and partial width for the $\Upsilon(4S)\rightarrow \Upsilon(1S)\pi^+\pi^-$ transition are shown in Table~\ref{tab:width} together with a summary of previously measured total and partial widths of the $\Upsilon$ resonances~\cite{pdgupsilon}.

\begin{table}[htb]
\center
\caption[]{\small  Total and partial widths of the $\Upsilon$ resonances (the subscript $\pi\pi$ refers to transitions $\Upsilon(nS)\rightarrow \Upsilon(1S)\pi^+\pi^-$). }\label{tab:width} 
\begin{tabular}{cccccc}
 Resonance & $\Gamma_{total}$ (keV) & $B_{ee}$ (\%) & $\Gamma_{ee}$ (keV) & $B_{\pi\pi}$ (\%) & $\Gamma_{\pi\pi}$ (keV) \\ \hline
 $\Upsilon(4S)$ & $(21\pm 4)\times 10^3$ & $(2.8\pm 0.7)\times 10^{-3}$ & $0.25\pm 0.03$ & $<0.012$ & $<  2.5$ \\
 \Uthree & $26.3\pm 3.5$ & $1.81\pm 0.17$ & $0.48\pm 0.08$ & $4.5\pm 0.2$ & $1.18\pm 0.17$ \\
 \Utwo  & $44\pm 7$ & $1.31\pm 0.21$ & $0.52\pm 0.03$ & $18.5\pm 0.8$ & $8.14\pm 1.34$\\
 \Uone  & $52.5\pm 1.8$ & $2.52\pm 0.17$ & $1.32\pm 0.05$ & --- & --- \\
\end{tabular}
\end{table}

\section{Measurement of the cross-sections \boldmath{\proeethreesgam} and \boldmath{\proeetwosgam}}

\subsection{Extraction of cross-sections}

\begin{figure}[htb]
\center
\epsfig{file=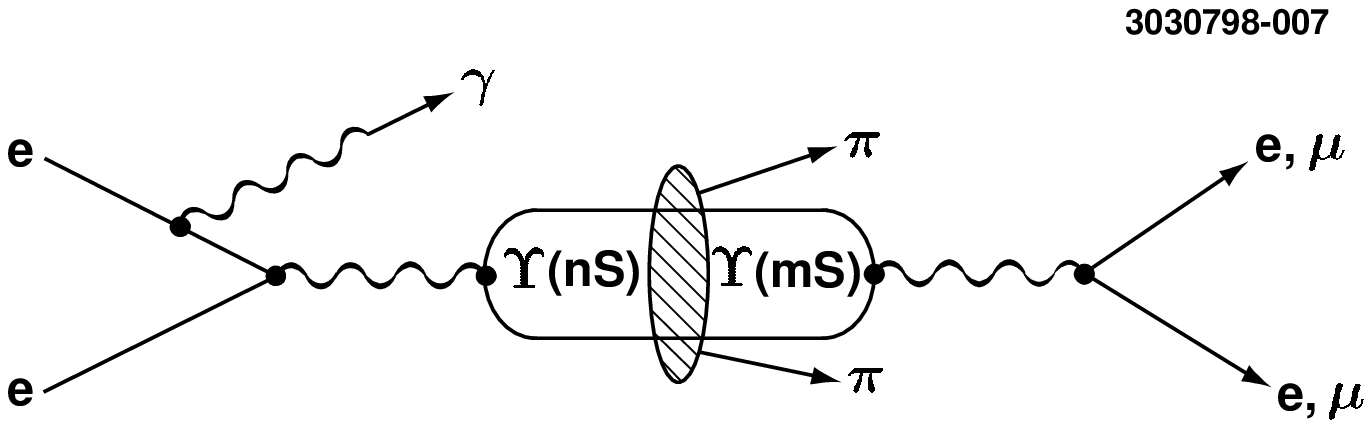}
\caption[]{Diagram for radiative production of the $\Upsilon$ resonances with subsequent dipion transitions.}
\label{fig:upsradprod}
\end{figure}

The same set of selection criteria used in our study of \Ufour\ dipion transitions was used in reconstruction of $\Upsilon(nS)$ radiative production events $e^+e^-\rightarrow \Upsilon(nS)\gamma$, $\Upsilon(nS)\rightarrow \Upsilon(mS)\pi^+\pi^-$, $\Upsilon(mS)\rightarrow e^+e^-,\mu^+\mu^-$ (Figure~\ref{fig:upsradprod}). Generally, we do not observe the initial state radiation photon; its presence is inferred from the shift of the observed $M_{recoil}$ peaks from the mass values of the corresponding $\Upsilon(mS)$. Because of the narrowness of the $\Upsilon(nS)$ resonances, the photons can be considered monochromatic. Effects due to the long Breit-Wigner tails of the $\Upsilon$ resonances are also included in our Monte Carlo efficiencies.

\begin{figure}[htb]
\center
\epsfig{file=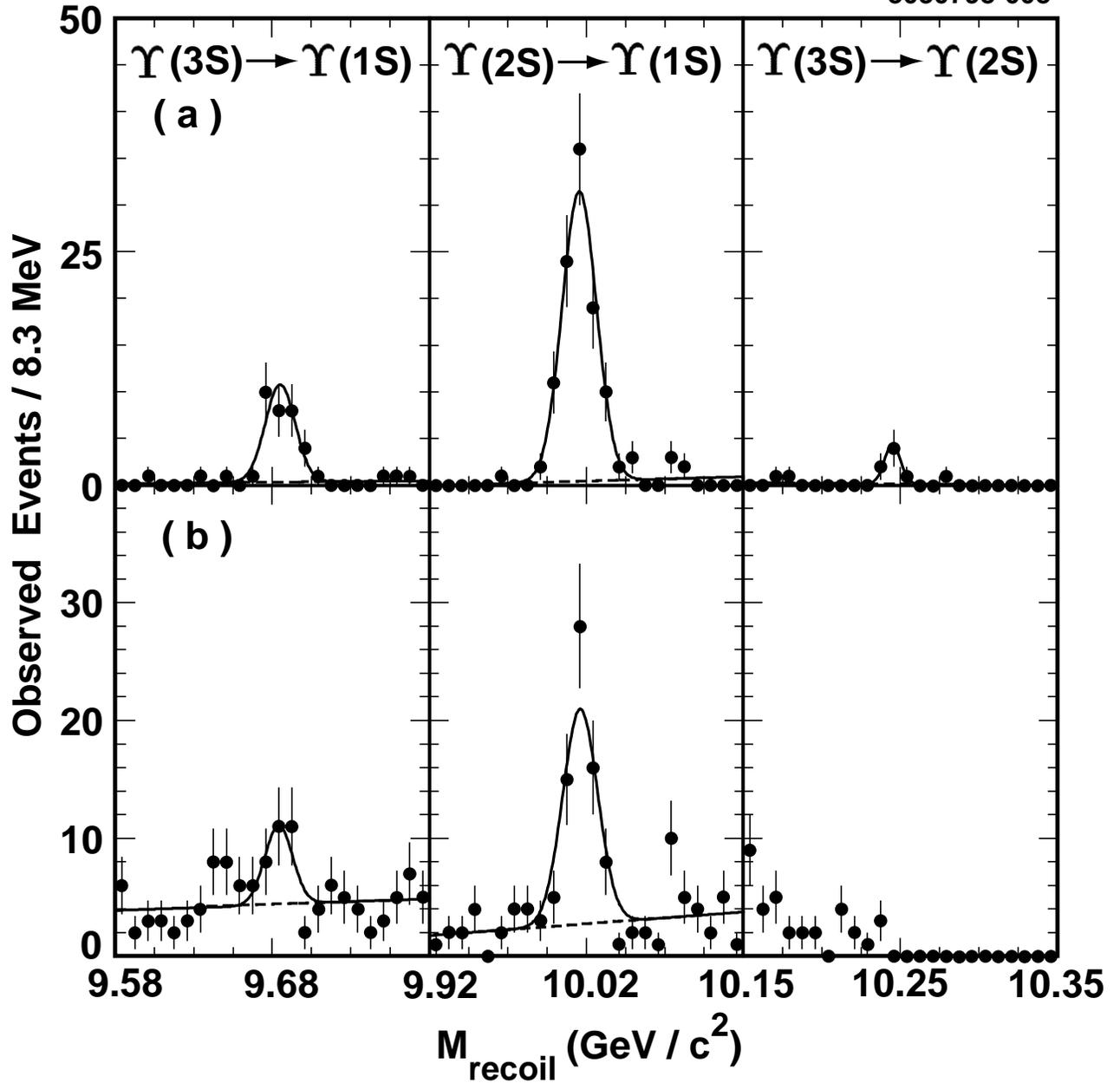, width=17cm}
\caption[]{Fits to the recoil mass distributions for the transitions \decnsmspipi\ at $\Ecm=10.58$ GeV for a) $\mu\mu$ channel and b) $ee$ channel.}\label{fig:fitson4s}
\end{figure}

\begin{figure}[htb]
\center
\epsfig{file=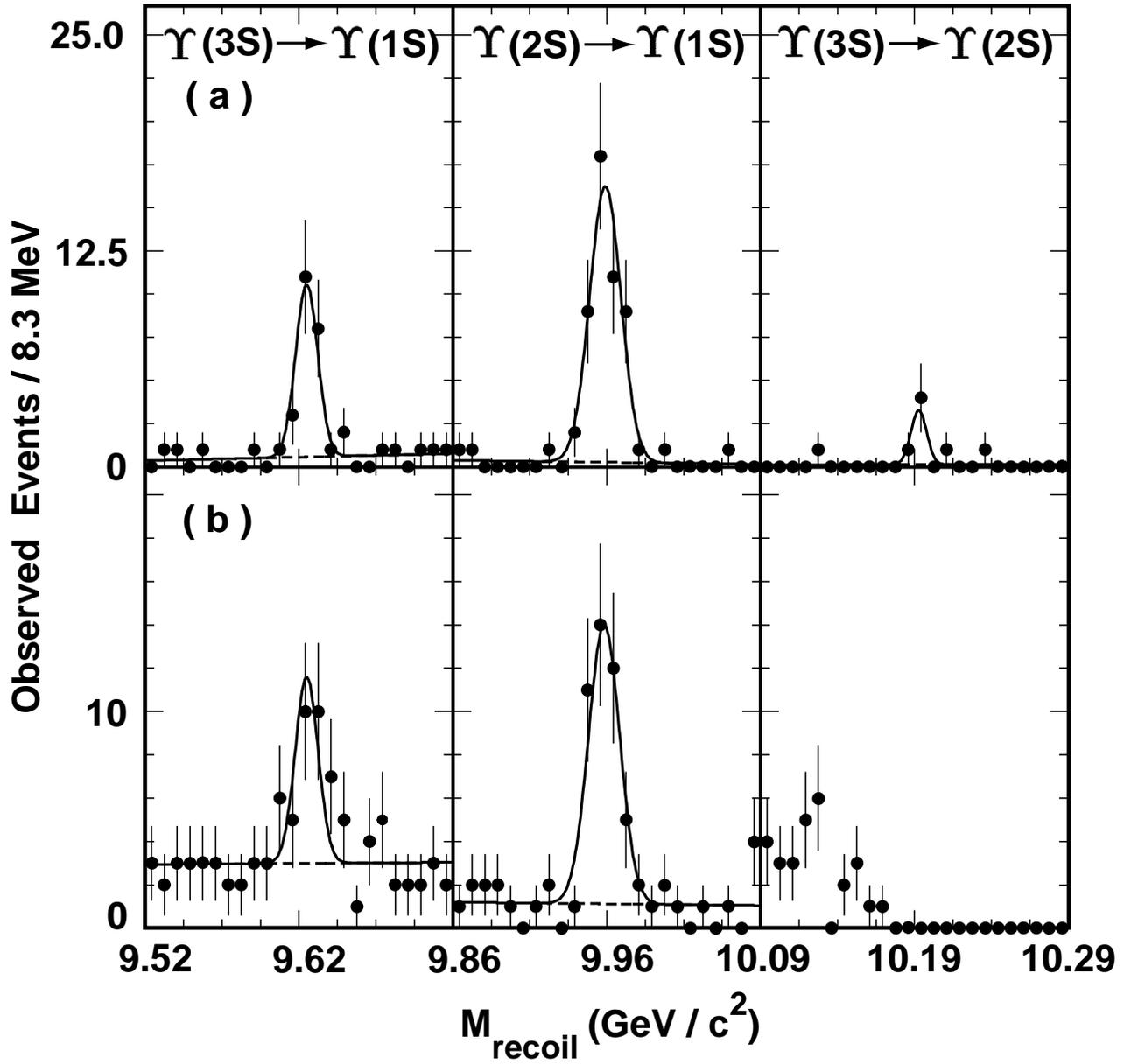, width=17cm}
\caption[]{Fits to the recoil mass distributions for the transitions \decnsmspipi\ at $\Ecm=10.52$ GeV for a) $\mu\mu$ channel and b) $ee$ channel.}\label{fig:fitsoff4s}
\end{figure}

We obtain the number of signal $\Upsilon(nS)$ radiative production events by fitting the recoil mass distributions corresponding to the data points inside our signal regions in Figure~\ref{fig:g2pi2ldtscat}. In Figure~\ref{fig:fitson4s} the $M_{recoil}$ distributions from data taken at $\Ecm=10.58$ GeV are shown; the same distributions from the data taken at $\Ecm=10.52$ GeV are shown in Figure~\ref{fig:fitsoff4s}. We see clear signals in the $\mu\mu$ channel for all three transitions of interest. In the $ee$ channel only the transition \dectwosonespipi\ has a clear peak while the transition \decthreesonespipi\ shows high background and the transition \decthreestwospipi\ is not seen at all because of the $M_{recoil}$ cutoff mentioned earlier. As a fitting function we use a Gaussian for the signal, plus a linear function to represent the background. In all cases the Gaussian width is fixed at the value from the corresponding fit of the Monte Carlo signal. As a check, we have also allowed the Gaussian widths to float in the $\mu\mu$ channel and used those widths to fit the $ee$ channel, obtaining consistent results.

\begin{table}[htb]
\center
\caption[]{\small  Efficiencies, yields, $C.L.$ of fit, and cross-sections for $e^+e^-\rightarrow \Upsilon(nS)\gamma$ at $\Ecm=10.58$ GeV.}\label{tab:rtnumberson4s}
\begin{tabular}{cccccc}
Transition & channel & $\epsilon$ (\%) & $N^{yield}$ & $C.L.$ (\%) & $\sigma$ (pb) \\ \hline
 $\Upsilon(3S)\rightarrow \Upsilon(1S)$ & $\mu\mu$ & $50.6\pm  2.0$ & $ 29.8\pm  5.7$ &  75.8  & $ 19.3\pm  3.7\pm   1.8$ \\
       & $ee$ & $40.1\pm  1.6$  & $ 18.1\pm   6.3$ &  58.5  & $ 14.5\pm   5.1\pm   1.7  $\\ \hline
 $\Upsilon(2S)\rightarrow \Upsilon(1S)$    & $\mu\mu$ & $46.9\pm  1.8$ & $102.1\pm  10.4$ &  17.1  & $ 17.3\pm  1.8\pm  1.5$\\
    & $ee$ & $35.7\pm  1.5$ & $ 61.1\pm   8.9$ &   3.0  & $ 13.4\pm  2.0\pm  1.4 $\\ \hline
 $\Upsilon(3S)\rightarrow \Upsilon(2S)$    & $\mu\mu$ & $14.8\pm  1.4$ & $  6.5\pm   2.7$ &  78.6  & $ 43.7\pm 17.8\pm  12.3$\\
  & $ee$ & $ 8.6\pm  0.9$  & --- & --- & --- \\
\end{tabular}
\end{table}

\begin{table}[hbt]
\center
\caption[]{\small  Efficiencies, yields, $C.L.$ of fit, and cross-sections for $e^+e^-\rightarrow \Upsilon(nS)\gamma$ at $\Ecm=10.52$ GeV.}\label{tab:rtnumbersoff4s}
\begin{tabular}{cccccc}
Transition & channel & $\epsilon$ (\%) & $N^{yield}$ &  $C.L.$ (\%) & $\sigma$ (pb) \\ \hline
  $\Upsilon(3S)\rightarrow \Upsilon(1S)$    & $\mu\mu$ & $49.6\pm  1.9$ & $ 21.5\pm  4.9$  &  89.5  & $ 26.8\pm  6.1\pm  2.6 $\\
  & $ee$ & $41.4\pm  1.7$ & $ 19.2\pm  5.9$  &  91.6  & $ 28.2\pm  8.6\pm  3.2$ \\ \hline
  $\Upsilon(2S)\rightarrow \Upsilon(1S)$    & $\mu\mu$ & $43.7\pm  1.7$ & $ 48.6\pm  7.1$  &  70.9  & $ 16.7\pm  2.4\pm  1.4 $\\
  & $ee$ & $36.5\pm  1.5$ & $ 39.1\pm  6.8$ &  47.0  &  $ 15.8\pm  2.8\pm   1.7$\\ \hline
 $\Upsilon(3S)\rightarrow \Upsilon(2S)$      & $\mu\mu$ & $15.2\pm  1.4$ & $  4.8\pm  2.3$ &  46.9  & $ 58.9\pm 27.9\pm 16.5$ \\
   & $ee$ & $ 9.2\pm  0.9$ & --- & --- & --- \\
\end{tabular}
\end{table}

In Tables~\ref{tab:rtnumberson4s} and~\ref{tab:rtnumbersoff4s} we report the efficiencies (obtained from a Monte Carlo simulation), yields, confidence levels of fits, and calculated cross-sections for the processes $e^+e^-\rightarrow \Upsilon(nS)\gamma$, using the formula $\sigma=N^{yield}/\epsilon {\cal L B}_{\pi\pi}{\cal B}_{ll}$. The numbers in Table~\ref{tab:rtnumberson4s} are for $\Ecm=10.58$ GeV data, and those in Table~\ref{tab:rtnumbersoff4s} are for $\Ecm=10.52$ GeV data. Relevant branching fractions are taken from the Particle Data Group~\cite{pdgupsilon}. The luminosity is ${\cal L}=\lumon$ fb$^{-1}$ for our $\Ecm=10.58$ GeV data sample and ${\cal L}=\lumoff$ fb$^{-1}$ for our $\Ecm=10.52$ GeV data sample. Appropriately combining the results from both dilepton channels we obtain the following averages for the cross-sections for $e^+e^- \rightarrow \Upsilon(nS)\gamma$ (we do not include $\Upsilon(3S)\rightarrow \Upsilon(2S)\pi\pi$  data in the averages):
\begin{itemize}
\item at $\Ecm=10.58$ GeV:
\[ \sigma(e^+e^- \rightarrow \Upsilon(3S)\gamma)=\xsecuthreeon\ \text{pb}\]
\[ \sigma(e^+e^- \rightarrow \Upsilon(2S)\gamma)=\xsecutwoon\ \text{pb} \] 
\item at $\Ecm=10.52$ GeV:
\[ \sigma(e^+e^- \rightarrow \Upsilon(3S)\gamma)=\xsecuthreeof\ \text{pb} \]
\[ \sigma(e^+e^- \rightarrow \Upsilon(2S)\gamma)=\xsecutwoof\ \text{pb} \]
\end{itemize}

\clearpage

\subsection{Systematic errors}

The sources and magnitudes of systematic errors are very similar to those in our \Ufour\ dipion transitions measurement. Two large additional errors are the uncertainties in the dipion branching ratios and the uncertainties in the shape of the fitting function (contributions from both the signal and the background functions are included). A summary of our systematic errors is given in Table~\ref{tab:rtsysterrors}. 

\begin{table}[htb]
\center
\caption[]{\small Sources and magnitudes of systematic errors for $e^+e^-\rightarrow\Upsilon(nS)\gamma$.}\label{tab:rtsysterrors}
\begin{tabular}{lccc}
        &\multicolumn{3}{c}{Systematic error (\%)} \\ \cline{2-4}
Source &\decthreesonespipi &\dectwosonespipi &\decthreestwospipi \\ \hline
Tracking                  & 2.8           & 2.8       & 8.5  \\ 
Finite MC sample          & 1.0           & 1.0       & 1.0  \\
Trigger efficiency        & 1.5           & 1.5       & 1.5 \\
Luminosity                & 0.9           & 0.9       & 0.9  \\
Fitting function          & 6.8           & 5.4       & 1.0  \\ 
Leptonic branching ratios & 6.7 / 2.8\tablenotemark[1] & 6.7 / 2.8\tablenotemark[1] & 16.0 \\
Dipion branching ratio    & 4.7           & 4.3       & 21.0 \\ \hline
Total                     & 11.2 / 9.4    & 10.2 / 9.2 & 27.8 \\
\end{tabular}
\tablenotetext[1]{separately for $ee/\mu\mu$ channels.}
\end{table}

\subsection{Discussion and conclusions}

\begin{figure}[hbt]
\center
\epsfig{file=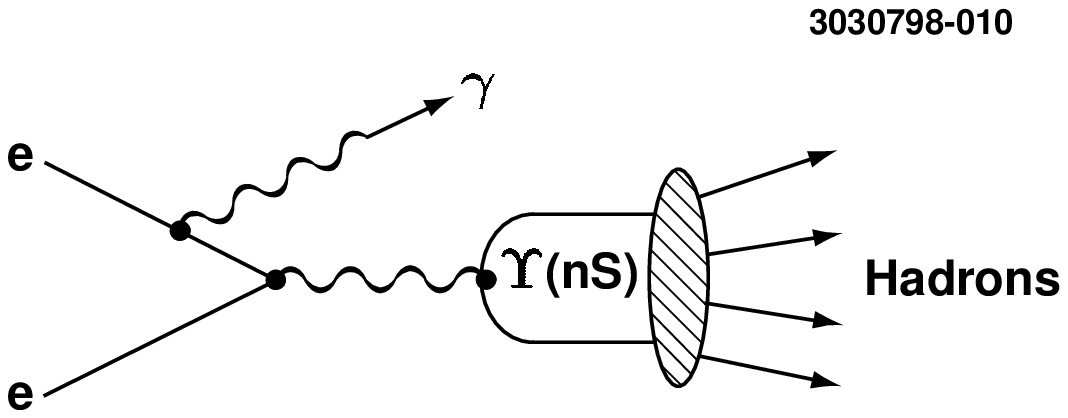}
\caption[]{Contributions from radiative production of the $\Upsilon$ resonances to the total hadronic cross-section $e^+e^-\rightarrow hadrons$.}
\label{fig:upsradprodall}
\end{figure}

Chetyrkin, K\"{u}hn, and Teubner (CKT)~\cite{Rtheory} performed a thorough calculation of the contributions to the total hadronic cross-section at $\Ecm=10.52$ GeV from the radiative production of the $\Upsilon$ resonances (Figure~\ref{fig:upsradprodall}). In Table~\ref{tab:predict} we compare our measurements of the $\Upsilon(nS)$ radiative production cross-sections with their predictions. Because we extract the total $e^+e^-\rightarrow \Upsilon(nS)\gamma$ cross-sections from our results (not just the hadronic part), numbers for the predicted cross-sections in Table~\ref{tab:predict} are scaled up by factors of 1.057, 1.041 and 1.081 for $\Upsilon(3S)$, $\Upsilon(2S)$ and $\Upsilon(1S)$, respectively, to account for the leptonic contribution. 

We note that similar ratios for the $\Upsilon(nS)$ radiative production cross-sections can be obtained from the following simple-minded arguments: (1) the initial state radiation photon spectrum varies as $dN/dE_{\gamma}\sim 1/E_{\gamma}$, and (2) the production of $\Upsilon$ resonances is proportional to their dielectron widths $\Gamma_{ee}$. Then one would expect for the production cross-section
\[ \sigma(e^+e^-\rightarrow \Upsilon\gamma)\propto \frac{\Gamma_{ee}}{E_{\gamma}}\]
At $\Ecm=10.58$ GeV this formula gives the ratio 2.3:1.0:1.3 for the radiative production cross-sections $\sigma^{\Upsilon(3S)}$:$\sigma^{\Upsilon(2S)}$:$\sigma^{\Upsilon(1S)}$, which is very close to the CKT predictions.

As seen in Table~\ref{tab:predict}, for the \Utwo\ case, the measured and CKT predicted values are in good agreement, while in the \Uthree\ case, the measured values are somewhat smaller than the predicted ones. Although the quoted systematic error in theoretical calculations is just a few permille, there may be large (on the level of 6-7 pb) shifts in the theoretical predictions due to uncertainties in the values of the $\Upsilon(3S)$ resonance parameters input to the theoretical model. Such shifts could easily reconcile our results with the CKT predictions.

\begin{table}[hbt]
\center
\caption[]{\small  Experimental and theoretical\tablenote{scaled up to include contribution from $\Upsilon(nS)$ leptonic modes.} values for the cross-sections of $e^+e^-\rightarrow \Upsilon(nS)\gamma$.}\label{tab:predict}
\begin{tabular}{ccccc}
        & \multicolumn{4}{c}{Cross-section (pb)} \\ \cline{2-5}
        & \multicolumn{2}{c}{$\Ecm=10.52$ GeV} & \multicolumn{2}{c}{$\Ecm=10.58$ GeV\tablenote{theoretical values for this energy are from~\cite{teubner}}}\\ \cline{2-3}\cline{4-5}
Process & Measured  & Predicted & Measured  & Predicted \\ \hline
 $e^+e^-\rightarrow \Upsilon(3S)\gamma$ & $\xsecuthreeof$ & 41.3 & $\xsecuthreeon$ & 30.7 \\
 $e^+e^-\rightarrow \Upsilon(2S)\gamma$ & $\xsecutwoof$   & 18.0 & $\xsecutwoon$ & 16.1 \\
 $e^+e^-\rightarrow \Upsilon(1S)\gamma$ & ---             & 20.4 & --- & 19.2\\ 
\end{tabular}
\end{table}

\section{Summary}

We have performed a search for the dipion transitions between the $\Upsilon$ resonances at the center of mass energies on and below the $\Upsilon(4S)$. We set 90\% confidence level upper limits on the branching fractions of the $\Upsilon(4S)$ dipion transitions: ${\cal B}(\Upsilon(4S)\rightarrow \Upsilon(2S)\pi^+\pi^-) < \ulfourstwos \times 10^{-4}$ and ${\cal B}(\Upsilon(4S)\rightarrow \Upsilon(1S)\pi^+\pi^-) < \ulfoursones \times 10^{-4}$. By observing the transitions $\Upsilon(3S)\rightarrow \Upsilon(1S)\pi^+\pi^-$ and $\Upsilon(2S)\rightarrow \Upsilon(1S)\pi^+\pi^-$, we have measured the cross-sections for the radiative processes $e^+e^- \rightarrow \Upsilon(3S)\gamma$ and $e^+e^- \rightarrow \Upsilon(2S)\gamma$. For the process $e^+e^- \rightarrow \Upsilon(2S)\gamma$ our results are in good agreement with theoretical predictions, while for the process $e^+e^- \rightarrow \Upsilon(3S)\gamma$ our measured values are somewhat below the predictions (Table~\ref{tab:predict}).

\acknowledgements

We gratefully acknowledge the effort of the CESR staff in providing us with
excellent luminosity and running conditions.
J.R. Patterson and I.P.J. Shipsey thank the NYI program of the NSF, 
M. Selen thanks the PFF program of the NSF, 
M. Selen and H. Yamamoto thank the OJI program of DOE, 
J.R. Patterson, K. Honscheid, M. Selen and V. Sharma 
thank the A.P. Sloan Foundation, 
M. Selen and V. Sharma thank Research Corporation, 
S. von Dombrowski thanks the Swiss National Science Foundation, 
and H. Schwarthoff thanks the Alexander von Humboldt Stiftung for support.  
This work was supported by the National Science Foundation, the
U.S. Department of Energy, and the Natural Sciences and Engineering Research 
Council of Canada.

\end{document}